\documentstyle[aps,epsfig]{revtex}
\begin{document}
\twocolumn[\hsize\textwidth\columnwidth\hsize\csname @twocolumnfalse\endcsname

\draft
\title{
Efficiency of Symmetric Targeting for Finite-$T$ DMRG}

\author
{ Tomotoshi Nishino \cite{Nishino} and Naokazu Shibata$^{1}$  }

\address{
Department of Physics, Graduate School of Science and
Technology, Kobe University, Rokkodai 657-8501\\
$^1$ Institute of Materials Science, University of Tsukuba,
Tsukuba 805-8578}


\maketitle
\begin{abstract}
Two targeting schemes have been known for the density matrix
renormalization group (DMRG) applied to non-Hermitian problems; one uses
an asymmetric density matrix and the other uses symmetric density
matrix. We compare the numerical efficiency of these two targeting
schemes when they are used for the finite temperature DMRG.
\end{abstract}

\vskip2pc]

\narrowtext



\section{Introduction}

The density matrix renormalization group (DMRG) established by
White~\cite{Wh1} has been successfully applied to various problems in
condensed matter physics.~\cite{Dresden} A recent technical
progress in DMRG is its applications to non-Hermitian problems, such as
asymmetric exclusion process,~\cite{ASEP1,ASEP2} reaction-diffusion
process,~\cite{DIF} and quantum Hall effect.~\cite{QHE}

For these non-Hermitian problems, left eigenvectors of the Hamiltonian
are not equal to the complex conjugates of the right eigenvectors.
Two different targeting schemes have been used for DMRG under
the situation. One is to use an asymmetric density matrix, which is a
partial trace between the left and the right eigenvectors.~\cite{Baxt}
This scheme has been used  for the DMRG applied to classical
systems~\cite{Nishi} and the finite temperature (finite-$T$)
DMRG.~\cite{Bur,Wang,Shib} The other scheme is to use a symmetric density
matrix, which is created by targeting both left and right
eigenvectors as two individual vectors.~\cite{Wh1,Dresden,DIF,QHE}

The purpose of this paper is to compare the numerical efficiencies of
these two schemes, by observing the cut-off error of the renormalization
group (RG) transformation applied to the finite temperature
Heisenberg spin chain. In the next section we define the cut-off errors as a
function of projection operator, that represents the freedom restriction
by the RG transformation. The two targeting schemes are briefly
reviewed in \S 3, and these schemes are compared by calculating the
cut-off error numerically. Conclusions are summarized in \S 4.

\section{Cut-off Error in RG Transformation}

The finite-$T$ DMRG~\cite{Bur,Wang,Shib} estimates the free energy of
one-dimensional quantum systems by way of a precise approximation for
the largest eigenvalue $\lambda$ of the quantum transfer
matrix~\cite{QTM} (QTM) ${\cal T}$. As an example of QTM, we consider
that of the $S = 1/2$ Heisenberg chain.
 (See Fig.1.) We express the
matrix element of the QTM as ${\cal T}_{i'_{~}\!\!j'_{~}\!\!,\,
ij }^{~}$, where $i$ $(i')$ and $j$ $(j')$ represents the upper-part
(U-part) and
the lower-part  (D-part~\cite{dhalf}) of the column spin, respectively.
It has been known that the
partition function of $N$-site system can be approximated by that of the
Trotter~\cite{Trot} decomposed two-dimensional classical system; $Z =
{\rm Tr} \, {\cal T}_{~}^N$. When $N$ is sufficiently large,  $Z$ is well
approximated as
\begin{equation}
{\cal T}_{~}^N \simeq {\bf V}_{~}^{\rm R}
\lambda_{~}^N \left( {\bf V}_{~}^{\rm L}  \right)_{~}^T \, ,
\end{equation}
where ${\bf V}^{\rm L}_{~}$ and ${\bf V}^{\rm R}_{~}$ is, respectively,
the left and the right eigenvector of ${\cal T}$, that satisfies the
eigenvalue relation
\begin{eqnarray}
\sum_{i'_{~}\!\!j'_{~}}^{~}
V^{\rm L}_{i'_{~}\!\!j'_{~}}
{\cal T}_{i'_{~}\!\!j'_{~}\!\!,\, ij }^{~}
&=&
V^{\rm L}_{ij} \lambda  \nonumber\\
\sum_{ij}^{~}
{\cal T}_{i'_{~}\!\!j'_{~}\!\!,\, ij }^{~}
V^{\rm R}_{ij}
&=&
\lambda V^{\rm R}_{i'_{~}\!\!j'_{~}} \, .
\end{eqnarray}
We have used the normalization
$\left( {\bf V}^{\rm L}_{~}, {\bf V}^{\rm R}_{~} \right)$
$= \sum_{ij}^{~} V_{ij}^{\rm L} V_{ij}^{\rm R}$ $= 1$ in Eq.1.
It should be noted that ${\bf V}^{\rm L}_{~}$ is not equal to ${\bf V}^{\rm
R}_{~}$ in general, because of the asymmetry ${\cal T} \neq {\cal T}_{~}^T$.

Let us consider a formal decomposition of these vectors into the
products of matrices
\begin{eqnarray}
V^{\rm L}_{ij} &\rightarrow& \sum_{\xi\eta}^{~}
O^{\rm L}_{i\xi} \, v^{\rm L}_{\xi\eta} \, Q^{\rm L}_{j\eta} \nonumber\\
V^{\rm R}_{ij} &\rightarrow& \sum_{\xi\eta}^{~}
O^{\rm R}_{i\xi} \, v^{\rm R}_{\xi\eta} \, Q^{\rm R}_{j\eta}
\end{eqnarray}
according to the convention in DMRG,~\cite{Dresden}
where $O^{\rm L/R}_{~}$ and $Q^{\rm
L/R}_{~}$ satisfy the orthogonal (or duality) relations
\begin{eqnarray}
\sum_i^{~} O^{\rm L}_{i\xi} O^{\rm R}_{i\xi'_{~}}
&=& \delta_{\xi\xi'_{~}}^{~} \nonumber\\
\sum_j^{~} Q^{\rm L}_{j\eta} Q^{\rm R}_{j\eta'_{~}}
&=& \delta_{\eta\eta'_{~}}^{~} \, ,
\end{eqnarray}
and $\xi$ and $\eta$ represent the block spin variables.
The matrices $O^{\rm L/R}_{~}$ and $Q^{\rm L/R}_{~}$ play the role of
renormalization group (RG) transformations, when the freedom
restriction $1 \leq \xi, \eta \leq m$ is considered for for both ${\bf
V}^{\rm L}_{~}$ and ${\bf V}^{\rm R}_{~}$ in Eq.3. Under the restriction,
$v^{\rm L}_{\xi\eta}$ and $v^{\rm R}_{\xi\eta}$ are $m$-dimensional
matrices that represent the renormalized states.

In DMRG applied to classical system or finite temperature quantum system,
the RG transformations $O^{\rm L/R}_{~}$ and $Q^{\rm
L/R}_{~}$ are determined  so that the cut-off error of the partition
function
\begin{figure}
\epsfxsize=72mm \epsffile{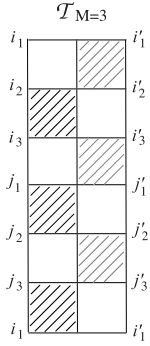}
\caption{
Quantum transfer matrix ${\cal T}_{i'_{~}\!\!j'_{~}\!\!,\, ij }^{~}$ of the
$S = 1/2$ Heisenberg spin chain, where the Trotter number $M$ is $3$,
and  both $i$ $(i'_{~})$ and $j$  $(j'_{~})$ consists of three $S = 1/2$ spin
variables.
}
\label{fig:1}
\end{figure}
\begin{equation}
\delta Z = {\rm Tr}  \, (1-P) {\cal T}_{~}^N =
\left( {\bf V}_{~}^{\rm L}\!, \, (1-P) {\bf V}_{~}^{\rm R} \right)
\lambda_{~}^N
\end{equation}
is suppressed, where $P$ is the projection operator
\begin{eqnarray}
P_{i'_{~}\!\!j'_{~}\!\!,\, ij }^{~} &=&
P_{i'_{~}\!i}^{\rm U} \, P_{j'_{~}\!j}^{\rm D} \nonumber\\
&=&
\sum_{\xi}^m O^{\rm R}_{i'_{~}\!\xi} O^{\rm L}_{i\xi} \,\,\,
\sum_{\eta}^m Q^{\rm R}_{j'_{~}\!\eta} Q^{\rm L}_{j\eta} \, ,
\end{eqnarray}
that represents the Hilbert space restriction by the RG transformation;
$P^{\rm U}_{~}$ and $P^{\rm D}_{~}$ is projection operator for U- and
D-part, respectively. Since the operator $1-P = 1 - P^{\rm U}_{~}P^{\rm
D}_{~}$ in Eq.5 can be factorized as
\begin{equation}
(1-P^{\rm U}_{~}) + (1-P^{\rm D}_{~}) -
(1-P^{\rm U}_{~})(1-P^{\rm D}_{~}) \, ,
\end{equation}
and the third term  is negligible
when it is applied to ${\bf V}_{~}^{\rm L}$ and
${\bf V}_{~}^{\rm R}$, we can precisely estimate
the relative cut-off error $\delta Z / Z$ by calculating the inner product
\begin{eqnarray}
&&
\left( {\bf V}_{~}^{\rm L}\!, \, (1-P^{\rm U}_{~}) {\bf V}_{~}^{\rm R} \right) +
\left( {\bf V}_{~}^{\rm L}\!, \, (1-P^{\rm D}_{~}) {\bf V}_{~}^{\rm R} \right)
\nonumber\\
&&
~~~~~~~ =
\left( 1 - {\rm Tr} \, P^{\rm U}_{~}\rho^{\rm U}_{~} \right) +
\left(1 - {\rm Tr} \, P^{\rm D}_{~}\rho^{\rm D}_{~} \right) \, ,
\end{eqnarray}
where $\rho^{\rm U}_{~}$ and $\rho^{\rm D}_{~}$ are the {\it asymmetric
density matrices}
\begin{eqnarray}
\rho^{\rm U}_{i'_{~}\!i}
&=& \sum_{j}^{~} V^{\rm L}_{i'_{~}\!j} V^{\rm R}_{ij}
\nonumber\\
\rho^{\rm D}_{j'_{~}\!j}
&=& \sum_{i}^{~} V^{\rm L}_{ij'} V^{\rm R}_{ij}
\end{eqnarray}
that satisfy the normalization ${\rm Tr} \rho^{\rm U}_{~}$ $= {\rm Tr}
\rho^{\rm D}_{~}$  $= \left( {\bf V}^{\rm L}_{~}, {\bf V}^{\rm R}_{~} \right)$
$= 1$. Let us keep in mind that ${\rm Tr} \, P^{\rm U}_{~}\rho^{\rm U}_{~}$
and ${\rm Tr} \, P^{\rm D}_{~}\rho^{\rm D}_{~}$ are essential for the cut-off
error of the RG transformation by $P = P^{\rm U}_{~}P^{\rm D}_{~}$.

\section{Asymmetric and Symmetric Targeting}

Two different targeting scheme have been used to determine the RG
transformation matrices $O^{\rm L/R}_{~}$ and $Q^{\rm L/R}_{~}$.
One is to obtain them by diagonalizing the asymmetric density
matrices~\cite{Baxt,Nishi,Wang,Shib} in Eq.9
\begin{eqnarray}
\rho^{\rm U}_{i'_{~}\!i} &\rightarrow& \sum_{\xi}^{~}
O^{\rm R}_{i'_{~}\!\xi} w^{~}_{\xi} O^{\rm L}_{i\xi}
\nonumber\\
\rho^{\rm D}_{j'_{~}\!j} &\rightarrow& \sum_{\eta}^{~}
Q^{\rm R}_{j'_{~}\!\eta} w^{~}_{\eta} Q^{\rm L}_{j\eta} \, ,
\end{eqnarray}
where $w^{~}_{\xi}$ is the common eigenvalue for both $\rho^{\rm U}_{~}$
and $\rho^{\rm D}_{~}$ in the order of decreasing absolute value; normally
all the $w^{~}_{\xi}$ are positive. The projection operators
$P^{\rm U}_{~}$, $P^{\rm D}_{~}$, and  $P$ created from $O^{\rm
L/R}_{~}$ and $Q^{\rm L/R}_{~}$ in Eq.10 are asymmetric, as
$\rho^{\rm U}_{~}$ and $\rho^{\rm D}_{~}$ are. Let us call such a
construction of $P^{\rm U/D}_{~}$ as {\it symmetric targeting.}
In this case, the relative cut-off error in Eq.8 can be calculated from the
eigenvalues of the asymmetric density matrix as
\begin{equation}
2 - {\rm Tr} \, P^{\rm U}_{~} \rho^{\rm U}_{~}
- {\rm Tr} \, P^{\rm D}_{~} \rho^{\rm D}_{~}
= 2(1 - \sum_{\xi}^m w^{~}_{\xi} ) \, .
\end{equation}
It is possible to choose $O^{\rm L/R}_{i\xi}$ and $Q^{\rm L/R}_{i\eta}$
so that $v^{\rm L}_{\xi\eta}$ and $v^{\rm R}_{\xi\eta}$ become
simultaneously diagonal: $v^{\rm L}_{\xi\eta}$ $= v^{\rm R}_{\xi\eta}$
$= \delta_{\xi\eta}^{~} \omega^{~}_{\xi}$ where $\omega^2_{\xi} =
w^{~}_{\xi}$. Thus we can interpret the decomposition in Eq.3
as an extension of the singular value decomposition for the dual vectors
${\bf V}^{\rm L}_{~}$ and ${\bf V}^{\rm R}_{~}$.~\cite{Nishi}

The other targeting scheme is to treat ${\bf V}^{\rm L}_{~}$ and ${\bf
V}^{\rm R}_{~}$ as individual vectors,~\cite{DIF,QHE}
as they simultaneously
target ground and excited states in DMRG applied to Hermitian
quantum systems.~\cite{Wh1}
In this case, the RG transformations are obtained by
first creating the symmetric density matrices
\begin{eqnarray}
{\bar \rho}^{\rm U}_{i'_{~}\!i} &=&
\frac{1}{2} \sum_{j}^{~} V^{\rm L}_{i'_{~}\!j} V^{\rm L}_{ij} +
\frac{1}{2} \sum_{j}^{~} V^{\rm R}_{i'_{~}\!j} V^{\rm R}_{ij}
\nonumber\\
{\bar \rho}^{\rm D}_{j'_{~}\!j} &=&
\frac{1}{2} \sum_{i}^{~} V^{\rm L}_{ij'} V^{\rm L}_{ij} +
\frac{1}{2} \sum_{i}^{~} V^{\rm R}_{ij'} V^{\rm R}_{ij}
\end{eqnarray}
and then by diagonalizing them
\begin{eqnarray}
{\bar \rho}^{\rm U}_{i'_{~}\!i} &\rightarrow& \sum_{\xi}^{~}
O^{~}_{i'_{~}\!\xi} {\bar w}^{\rm U}_{\xi} O^{~}_{i\xi}
\nonumber\\
{\bar \rho}^{\rm D}_{j'_{~}\!j} &\rightarrow& \sum_{\eta}^{~}
Q^{~}_{j'_{~}\!\eta} {\bar w}^{\rm D}_{\eta} Q^{~}_{j\eta} \, ,
\end{eqnarray}
where we have dropped the label ${\rm L}$ and ${\rm R}$ from
$O^{\rm L/R}_{~}$ and $Q^{\rm L/R}_{~}$, because
$O^{\rm L}_{~} = O^{\rm R}_{~}$ and $Q^{\rm L}_{~} = Q^{\rm R}_{~}$.
In this case, the projection operators
\begin{eqnarray}
{\bar P}^{\rm U}_{i'_{~}\!i} &=&
\sum_{\xi}^m O^{~}_{i'_{~}\!\xi} O^{~}_{i\xi}
\nonumber\\
{\bar P}^{\rm D}_{j'_{~}\!j} &=&
\sum_{\eta}^m Q^{~}_{j'_{~}\!\eta} Q^{~}_{j\eta}
\end{eqnarray}
are symmetric. Let us call such a construction of ${\bar P}^{\rm U/D}_{~}$
as {\it symmetric targeting.} Unlike the asymmetric targetting,
it is impossible to make the $m$-dimensional
matrices $v^{\rm L}_{\xi\eta}$ and  $v^{\rm R}_{\xi\eta}$ simultaneously
diagonal. This targeting scheme is often used because there is no
need to diagonalize the asymmetric density matrix, which requires special
numerical  care.~\cite{reortho} It should be noted that the relative cut-off
error
\begin{equation}
\delta Z / Z =
1 - {\rm Tr} \, {\bar P}^{\rm U}_{~} \rho^{\rm U}_{~} +
1 - {\rm Tr} \, {\bar P}^{\rm D}_{~} \rho^{\rm D}_{~}
\end{equation}
in the symmetric targeting is not directly related to the eigenvalues of the
symmetric density matrices in Eq. 12.

Now let us compare the relative cut-off errors for both asymmetric and
symmetric targeting. We choose the $S = 1/2$ isotropic Heisenberg spin chain
as the reference system, and
fix the Trotter number $M = 7$ so that we can obtain
the eigenvector of ${\cal T}$; we have to know
${\bf V}_{~}^{\rm L}$ and ${\bf V}_{~}^{\rm R}$ exactly
in order to evaluate $\delta Z / Z$.
We consider the case where ${\rm
U}$-part contains the same number of spin variables as ${\rm D}$-part;
this U-D division is normally used for the infinite
system algorithm.~\cite{Wh1}
Since ${\rm U}$-part is identical to the ${\rm D}$-part,
$1 - {\rm Tr} \, P^{\rm U}_{~} \rho^{\rm U}_{~}$
$= 1 - {\rm Tr} \, P^{\rm D}_{~} \rho^{\rm D}_{~}$
holds for Eq.11 and
$1 - {\rm Tr} \, {\bar P}^{\rm U}_{~} \rho^{\rm U}_{~}$
$= 1 - {\rm Tr} \, {\bar P}^{\rm D}_{~} \rho^{\rm D}_{~}$
for Eq.15.  Figure 2 shows the relative cut-off errors
%
%
when the imaginary time step $J \Delta \tau $~\cite{Shib} is equal to $1/7$.
As it is seen, $1 - {\rm Tr} \, P^{\rm U}_{~} \rho^{\rm U}_{~}$ decreases
monotonically with respect to $m$, and is always positive. On the other
hand,  the dumping of $1 - {\rm Tr} \, {\bar P}^{\rm U}_{~} \rho^{\rm U}_{~}$
with respect to $m$ is oscillatory;
$1 - {\rm Tr} \, {\bar P}^{\rm U}_{~} \rho^{\rm U}_{~}$
is not always positive,~\cite{nonpos} and the calculated partition function
is not the variational lower bound. For most of $m$ the error
$1 - {\rm Tr} \, P^{\rm U}_{~} \rho^{\rm U}_{~}$ is smaller than
$| 1 - {\rm Tr} \, {\bar P}^{\rm U}_{~} \rho^{\rm U}_{~} |$, that
shows the superiority of asymmetric targeting for the
finite-$T$ DMRG.

\begin{figure}
\epsfxsize=72mm \epsffile{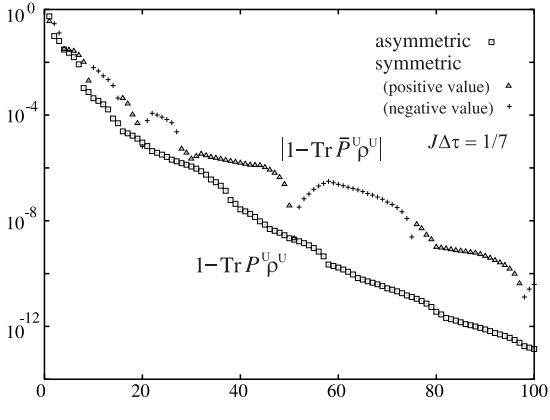}
\caption{
The relative cut-off error for the asymmetric targeting
$1 - {\rm Tr} \, P^{\rm U}_{~} \rho^{\rm U}_{~}$ in Eq.11 and that
for the symmetric targeting
$1 - {\rm Tr} \, {\bar P}^{\rm U}_{~} \rho^{\rm U}_{~}$ in Eq.15
when $J \Delta \tau = 1/7$.
For the latter we use triangle mark when it is positive,
and use cross mark when negative.
The horizontal axis shows the number of states kept.
}
\label{fig:2}
\end{figure}

\begin{figure}
\epsfxsize=72mm \epsffile{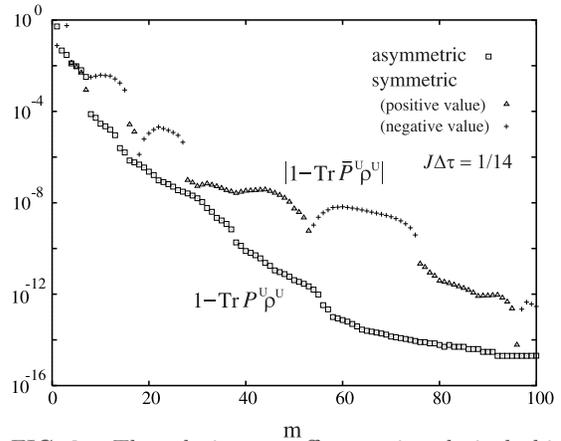}
\caption{
The relative cut-off errors in relatively high temperature
$J\Delta\tau = 1/14$.
}
\label{fig:3}
\end{figure}

Figure 3 shows the cut-off errors for relatively
high temperature $J \Delta \tau = 1/14$.
In both Figs.2 and 3, we have to keep twice as large as $m$ for the
symmetric targeting in order to keep the same
cut-off error of the asymmetric targeting. This may be explained
by the fact that the asymmetric projection operator
$P^{\rm U/D}_{~}$ is created by $2m$
numbers of linearly independent vectors, while
the symmetric projection operator ${\bar P}^{\rm U/D}_{~}$
is created by $m$ numbers of orthogonal vectors.

\section{Conclusion}

We have compared the numerical efficiency of the symmetric and
asymmetric targeting schemes when they are applied to the finite
temperature DMRG. It is shown that the cut-off error calculated by
the symmetric targeting is larger than that of asymmetric targeting;
as far as cut-off error is concerned, the asymmetric targetting is
superior to the symmetric targetting.

If we keep twice as large as $m$ for the symmetric targeting,
we can recover the numerical precision of the asymmetric targeting.
Therefore, for the problems that does not require large $m$ for the DMRG
calculations, the
symmetric targeting is of use, in the sense that it does not require
the diagonalization of asymmetric density matrix, and is free from
complex eigenvalue problem.~\cite{Shib,reortho}

T.N. thank to  G.~Sierra, M.~A.~Mart\'{\i}n-Delgado, and S.~R.~White
for the discussion about multi state targeting. The
present work is partially supported by a Grant-in-Aid from Ministry of
Education, Science and Culture of Japan.


\begin{thebibliography}{99}
\bibitem{dhalf} We use `D-part' instead of `L-part', because we will use
     the label L and R in order to denote left and right, respectively.
\bibitem{nonpos} Note that the density matrix $\rho^{\rm U}_{~}$
is asymmetric, and ${\bar P}^{\rm U}_{~}$ is the symmetric projection
operator. In such a case,
$1 \geq {\rm Tr} \, {\bar P}^{\rm U}_{~} \rho^{\rm U}_{~}$
is not always true.
\end{thebibliography}
\end{document}